\documentclass[twocolumn,showpacs,amsmath,amssymb,superscriptadress]{revtex4}

\usepackage{dcolumn}

\usepackage{graphicx}
\usepackage{bm}

\begin{document}


\title
{Microscopic Formulation of Interacting Boson Model for Rotational Nuclei}

\author{Kosuke Nomura$^1$}
\author{Takaharu Otsuka$^{1,2,3}$}
\author{Noritaka Shimizu$^1$}
\author{Lu Guo$^{4}$}

\affiliation{$^1$Department of physics, University of Tokyo, Hongo,
Bunkyo-ku, Tokyo, 113-0033, Japan}
\affiliation{$^2$Center for Nuclear Study, University of Tokyo, Hongo,
Bunkyo-ku Tokyo, 113-0033, Japan} 
\affiliation{$^3$National Superconducting Cyclotron Laboratory, 
Michigan State University, East Lansing, MI}
\affiliation{$^4$RIKEN Nishina Center, Hirosawa, Wako-shi, Saitama
351-0198, Japan}  

\date{\today}

\begin{abstract}
We propose a novel formulation of the Interacting Boson Model (IBM) for
rotational nuclei with axially-symmetric strong deformation.  
The intrinsic structure represented by the potential energy surface 
(PES) of a given multi-nucleon system has a certain similarity 
to that of the corresponding multi-boson system.  Based on this feature, 
one can derive an appropriate boson Hamiltonian as already reported.
This prescription, however, has a major difficulty in rotational
spectra of strongly deformed nuclei: the bosonic moment of inertia 
is significantly smaller than the corresponding nucleonic one. 
We present that this difficulty originates in the difference between
the rotational response of a nucleon system and that of the 
corresponding boson system, and could arise even if the PESs of the two 
systems were identical. 
We further suggest that the problem can be cured by implementing 
$\hat{L} \cdot \hat{L}$ term into the IBM Hamiltonian, with  
coupling constant derived from the cranking 
approach of Skyrme mean-field models. 
The validity of the method is confirmed for rare-earth and actinoid 
nuclei, as their experimental rotational yrast bands are reproduced nicely.
\end{abstract}


\maketitle

The atomic nucleus is a strongly interacting many-body quantal system which 
has collective properties resulting in various deformed shapes.  
If a nucleus is strongly deformed, it rotates, exhibiting characteristic
rotational band structure with remarkable regularity.  Such rotational 
motion can be viewed as a manifestation of symmetry 
restoration mechanism of Nambu, and has attracted much attention 
in nuclear physics from various viewpoints \cite{BM,RS}. 
 
The Interacting Boson Model (IBM) \cite{IBM,IBMapp}
has been successful in phenomenological studies for describing 
low-lying quadrupole collective states of medium-heavy nuclei 
\cite{IBM}. 
The major assumption of the IBM is to employ $L=0^{+}$ ($s$) and 
$L=2^{+}$ ($d$) bosons which reflect the collective $S$ and $D$ pairs
of valence nucleons \cite{OAI}. 
The microscopic foundation of the IBM has been studied extensively 
so as to derive an IBM Hamiltonian starting from 
nucleon degrees of freedom \cite{OAI,MO,Odef}. 
A new approach to derive the IBM Hamiltonian has been
presented recently \cite{nso}. 
In this approach, the potential energy surface (PES) with quadrupole
degrees of freedom, obtained from the mean-field calculation with the
Skyrme Energy Density Functional (EDF) \cite{Sk,VB}, is compared to 
the corresponding PES of IBM to obtain the parameters of IBM Hamiltonian. 
As a Skyrme EDF gives universal descriptions of various nuclear 
properties \cite{RS,Sk,VB,Bender_review}, one can derive the IBM 
Hamiltonian basically for all situations in a unified way.
This method turned out to be valid particularly for nuclei
with weak to moderate quadrupole deformation, and has been practiced
extensively \cite{nsofull}. 
When a nucleus is well deformed, however, the nucleonic rotational
spectrum appears notably and systematically different from the 
corresponding bosonic one, being manifested by too small bosonic moment of
inertia as compared to the corresponding fermionic one
\cite{nso,nsofull}. 

This kind of difference has been known as a result of limited 
degrees of freedom, $s$ and $d$ bosons only, 
in many cases \cite{OtsukaPhD,schematic}.   
In order to remedy this problem, another type of nucleon
pairs, e.g., $L=4^{+}$ ($G$) pair, and the corresponding boson image
($g$ boson) have been introduced, and their effects were renormalized
into $sd$ boson sector, yielding IBM Hamiltonians 
consistent with phenomenological ones    
\cite{MO,Odef,OtsukaPhD,schematic,defSM,Scholten_OAI,Yoshinaga1984SDG,Zirnbauer1984MIBM,g-ren,Pannert1985HFBIBM,Otsuka1988sdgIBM}. 
In the mean time, the validity of IBM for rotational nuclei was
analyzed in terms of the Nilsson model \cite{BMc}, 
coming up with the criticism that the $SD$-pair truncation may be 
far from sufficient for describing intrinsic states of 
strongly deformed nuclei, and this naturally casts a question 
on the applicability of the IBM to rotational nuclei in particular. 
While it has been reported that the $SD$-pair dominance holds 
to a good extent in intrinsic states of rotational nuclei
\cite{defSM,OAY,gang4},    
there has been no conclusive mapping procedure from nucleonic
systems to IBM ones covering rotational nuclei.
It is thus of much interest to revisit this issue 
with the newly proposed method of Ref.~\cite{nso}, 
looking for a prescription to cure the afore-mentioned problem 
of too small moment of inertia. 

In the method of Ref.~\cite{nso}, 
we calculate the energies of nucleonic and bosonic intrinsic states 
representing various shapes, and obtain PESs.
We then determine parameters of the IBM Hamiltonian so that the bosonic 
PES becomes similar to the nucleonic one \cite{nso}.
These intrinsic states are at rest with 
rotational frequency $\omega$=0.
In this paper, we move on by one step further with non-zero rotational
frequency $\omega \neq$ 0.  Actually we analyze the responses of the
nucleonic and bosonic intrinsic states 
by rotational cranking with infinitesimal $\omega$.  
From such responses, one can extract the most important rotational 
correction to the IBM Hamiltonian.

\begin{figure}
 \includegraphics[width=8.0cm]{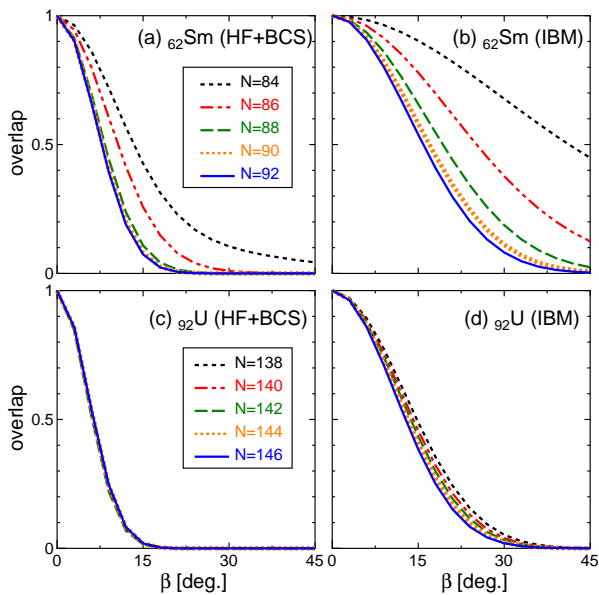}
 \caption{(Color online) Overlap between the intrinsic state and its
 rotation at angle $\beta$ for $^{146-154}$Sm and $^{230-238}$U nuclei
 for (a)(c) fermion (HF+BCS) and (b)(d) boson (IBM) systems. } 
\label{fig:overlap}
\end{figure}

The nucleon intrinsic state $|\phi_F\rangle$ is obtained from the 
Hartree-Fock plus BCS (HF+BCS) calculation. 
Skyrme SkM* interaction
\cite{SkM} is used throughout, while 
different Skyrme forces do not alter the conclusion. 

For boson system, we consider the IBM-2, because it is closer to a
microscopic picture than the simpler version of IBM. 
The IBM-2 is comprised of proton $L=0^+$
($s_{\pi}$) and $2^+$ ($d_{\pi}$) bosons, and of neutron $L=0^+$
($s_{\nu}$) and $2^+$ ($d_{\nu}$) bosons \cite{OAI}. 
We take the standard IBM-2 Hamiltonian,
\begin{eqnarray}
 H_{B}=\epsilon n_{d}+\kappa Q_{\pi}\cdot Q_{\nu}, 
\label{eq:bh0}
\end{eqnarray}  
where $n_{d}=n_{d\pi}+n_{d\nu}$ with $n_{d\pi}$ ($n_{d\pi}$) being the
proton (neutron) d-boson number operator and 
$Q_{\rho}=s_{\rho}^{\dagger}\tilde d_{\rho}+d_{\rho}^{\dagger}\tilde
s_{\rho} + \chi_{\rho}[d_{\rho}^{\dagger}\tilde d_{\rho}]^{(2)}$. 
Here $\epsilon$, $\kappa$, and $\chi_{\pi,\nu}$ are parameters, and
their values are determined by comparing nucleonic and bosonic 
PESs following Refs. \cite{nso,nsofull}.
The boson intrinsic state $|\phi_B\rangle$ is written in general 
as a coherent state \cite{coherent1,coherent2} 
\begin{eqnarray}
 |\phi_B\rangle\propto
\prod_{\rho=\pi,\nu}(s^{\dagger}_{\rho}+\sum_{\mu=\pm2,\pm1,0}a_{\rho\mu}d^{\dagger}_{\rho\mu})^{n_{\rho}}|0\rangle, 
\label{eq:coherent}
\end{eqnarray}
where $|0\rangle$ and $a_{\rho\mu}$ represent the boson vacuum (inert
core) and amplitude, respectively. 

We now look into the problem of rotational response.
We shall restrict ourselves to nuclei with axially symmetric 
strong deformation,
because this problem is crucial to those nuclei  
but is not so relevant to the others.
An axially symmetric intrinsic state is invariant with respect to the 
rotation around the symmetry ($z$) axis.  
This means $a_{\rho\mu}=0$ for $\mu \neq 0$ in eq.~(\ref{eq:coherent})
in the case of bosons.  
Such intrinsic states of nucleons and bosons are supposed to be obtained 
as the minima of the PESs.
Let us now rotate the axially symmetric intrinsic states about the 
$y$-axis by angle $\beta$. 
Figure~\ref{fig:overlap} shows the overlap between the intrinsic state 
$|\phi_X\rangle$ and the rotated one 
$|\phi_X^{\prime}\rangle=e^{-iL_y\beta}|\phi_X\rangle$, 
where $X$ stands for either fermion ($X$=$F$) or boson ($X$=$B$). 
Here $L_y$ denote the $y$-component of the angular momentum operator.
We take $^{146-154}$Sm and $^{230-238}$U as examples.  
Some of these nuclei are good examples of SU(3) limit of IBM \cite{SU3}. 

Figures \ref{fig:overlap}(a,c) and \ref{fig:overlap}(b,d) show  
the overlaps for nucleons and bosons, respectively.
For Sm isotopes, the parameters of $H_B$ are taken from \cite{nsofull}, 
while the parameters for U isotopes are determined as 
$\epsilon\approx 0.100$ MeV, $\kappa\approx -0.18$ MeV, and
$\chi_{\pi}\approx\chi_{\nu}\approx-1.0$. These parameters are 
used throughout this paper.

In each case, the overlap is peaked at $\beta=0^{\circ}$ 
with the value unity, and decreases with $\beta$.
The nucleonic overlaps are peaked more sharply, whereas boson ones 
are damped more slowly.  It is clear that 
as a function of $\beta$, boson rotated intrinsic state changes
more slowly than the corresponding nucleon one, due to limited degrees 
of freedom for bosons. 

We point out that the overlap becomes narrower in $\beta$ with the 
neutron number $N$ for Sm isotopes (see Fig.~\ref{fig:overlap}(a,b)). 
This is related to the growth of deformation. 
On the other hand, there is no notable change in the overlap for 
these U isotopes, because pronounced prolate minimum appears always 
at $\beta_{2}\sim 0.25$ in their PES.

The nucleon-boson difference of the rotational response 
discussed so far 
suggests that the rotational spectrum of a nucleonic system 
may not be fully reproduced by the boson system determined by 
the mapping method of Ref. \cite{nso} using the PESs at rest.
In fact, it will be shown later 
that the moment of inertia of a nucleon system differs from the one 
calculated by the mapped boson Hamiltonian.
We then introduce a term into the
boson Hamiltonian, so as to keep the PES-based mapping procedure 
but incorporate the different rotational responses.  
This term takes the form of $\hat{L}\cdot \hat{L}$ where $\hat{L}$
denotes the boson angular momentum operator.
This term is nothing but the squared magnitude of the angular
momentum with the eigenvalue $L(L+1)$, and changes the moment of 
inertia of rotational band keeping their wave functions.  
A phenomenological term of this form was used in the fitting  
calculation of IBM, particularly in its SU(3) limit \cite{SU3}, 
without knowing its origin or physical significance. 

We adopt, hereafter, a Hamiltonian, $H_{B}^{\prime}$, which includes 
this term with coupling constant $\alpha$ :  
\begin{eqnarray}
 H_{B}^{\prime}=H_{B}+\alpha \hat{L}\cdot \hat{L}, 
\label{eq:bh}
\end{eqnarray}
where $H_{B}$ is given in Eq.~(\ref{eq:bh0}). 
The $\alpha \hat{L}\cdot \hat{L}$ term will be referred to 
as LL term hereafter. The LL term contributes to the PES in the same 
way as a change of $d$-boson energy $\Delta \epsilon = 6 \alpha$ 
(see eq.~(\ref{eq:bh0})), because the PES at rest ({\it i.e.}, 
$\omega$=0) is formed by the boson intrinsic state $|\phi_B\rangle$  
containing no $d_{\pm 1}$ component. 
Hence, by shifting $\epsilon$ slightly, 
we obtain the same PES as the one without the LL term, 
and consequently the other parameters of mapped $H_{B}$ 
remain unchanged.

We now turn to the determination of $\alpha$ in Eq.~(\ref{eq:bh}). 
First, we perform the cranking model calculation for the fermion 
system to obtain its moment of inertia, denoted by 
${\cal J}_{F}$, in the usual way \cite{RS}. 
By taking the Inglis-Belyaev (IB) formula, we obtain 
\cite{Inglis,Belyaev} 
\begin{eqnarray}
{\cal J}_{F}
=2\cdot\sum_{i,j>0}
\frac{|\langle i|L_{k}|j\rangle|^2}
{E_{i}+E_{j}}
(u_{i}v_{j}-u_{j}v_{i})^2, 
\label{eq:IB}
\end{eqnarray}
where energy $E_{i}$ and v-factor $v_{i}$ of quasi-particle 
state $i$ are calculated by the HF+BCS method of Ref.~\cite{Guo}.
Here, $L_{k}$ is the nucleon angular momentum operator, and 
$k$ means the axis of the cranking rotation, being either $x$ or
$y$, as $z$-axis.  Based on the earlier argument, the $y$-axis
is chosen. 

Next, the bosonic moment of inertia, denoted as ${\cal J}_{B}$, 
is calculated by the cranking formula of 
Ref. \cite{IBMcranking} with $d_{\pm 1}$ being mixed,  
to an infinitesimal order, into the coherent state 
$|\phi_{B}\rangle$ in Eq.~(\ref{eq:coherent}): 
\begin{eqnarray}
 {\cal J}_{B}=\lim_{\omega\to 0}
\frac{1}{\omega}
\frac{\langle\phi_B|L_{k}|\phi_B\rangle}{\langle\phi_B|\phi_B\rangle}, 
\label{eq:IBM-MOI}
\end{eqnarray} 
where $\omega$ is the cranking frequency, $a_{\rho\pm 1}$ denotes 
the amplitude for $d_{\rho\pm 1}$, and 
$L_{k}$  stands for the boson angular momentum operator.  
Note that $a_{\rho\pm 1}$ $\propto$ $\omega$ at this limit, 
leading ${\cal J}_{B}$ to a finite value.  

The value of $\alpha$ is determined for individual
nucleus so that the corresponding bosonic 
moment of inertia, ${\cal J}_{B}$ in eq.~(\ref{eq:IBM-MOI}) 
becomes equal to ${\cal J}_{F}$ in eq.~(\ref{eq:IB}).
This prescription makes sense, 
if the nucleus is strongly deformed and the fixed intrinsic state
is so stable as to produce individual levels of a rotational band
through the angular momentum projection in a good approximation.  
The resultant excitation energies should follow the rotor formula
$E_x \propto L(L+1)$ for $L$ being the angular momentum of the level.
The present prescription with the LL term should be applied only to
certain nuclei which belong to this type.  We introduce a criterion
to select such nuclei in terms of the ratio 
$R_{4/2} \, = \, E_{\rm x}(4^{+}_{1})/E_{\rm x}(2^{+}_{1})$, 
and set a minimum value for this.  
Empirical systematics \cite{Cakirli2006pn} suggests that 
the evolution towards stronger deformation continues as the 
number of valence nucleons increases, but this evolution becomes 
saturated beyond $R_{4/2} \sim$ 3.2.  Namely, for the nuclei 
with $R_{4/2} >$ 3.2, the deformation is considered to be evolved 
sufficiently well, and we take $R_{4/2} >$ 3.2 as the criterion 
to apply the LL term.  
This discrete criterion is also for the sake of simplicity, but 
the major discussions of this work do not depend on its details.

\begin{figure}
 \includegraphics[width=8.0cm]{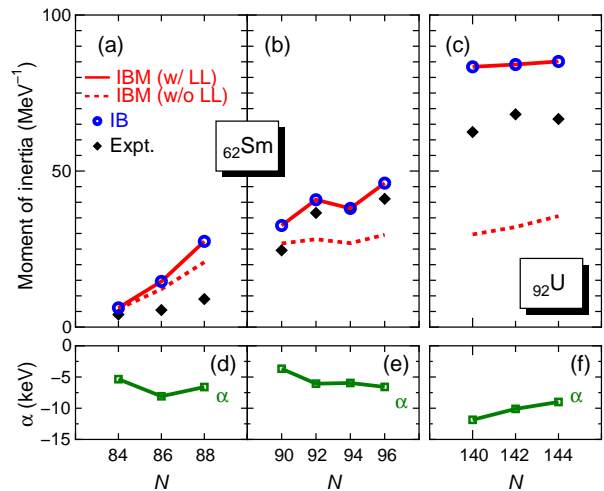}
 \caption{(Color online) 
(Upper panels) Moments of inertia in the intrinsic state for (a)
 $^{146-150}$Sm, (b) $^{152-158}$Sm and (c) $^{232-236}$U, calculated by
 IBM with (w/) and without (w/o) the LL  
 term and by IB formula. Experimental data taken from $E_{\rm x}(2^{+}_{1})$ \cite{data} are also shown. 
(Lower panel) The derived $\alpha$ value for (d) 
 $^{146-150}$Sm, (e) $^{152-158}$Sm and (f) $^{232-236}$U. }  
\label{fig:MOI}
\end{figure}

Figures~\ref{fig:MOI}(a) $\sim$ (c) show the
moments of inertia for Sm and U isotopes. 
In these figures, ${\cal J}_{B}$ calculated with the LL term (w/ LL), 
${\cal J}_{B}$ calculated without it (w/o LL), 
and ${\cal J}_{F}$ are compared.  
Experimental ones determined from the $2^{+}_{1}$ levels \cite{data} 
are shown also. 

We divide Sm isotopes into two categories according to the criterion
defined above.  First, the ratio $R_{4/2}$ is calculated without the LL 
term, leading to $^{152-158}$Sm with $R_{4/2}>3.2$ and $^{146-150}$Sm 
with $R_{4/2}<3.2$.  For the former category, the LL term should be 
included, and Fig.~\ref{fig:MOI}(b) demonstrates that    
the LL term produces significant effects so as to be
just enough for the agreement to experiment.  
To be more precise, 
the experimental value becomes large for $N=90$, and looks 
nearly flat for $N\geqslant 92$, being $35\sim 40$ MeV$^{-1}$.  
The enlargement of moment inertia means that the value of $\alpha$ 
is negative. 
The IB formula reproduces this trend quite well, which is 
inherited to bosons by the present method.    

Although the LL term should not be used for the category   
depicted in Fig.~\ref{fig:MOI}(a), we shall study some features.
The increase of ${\cal J}_{F}$ and ${\cal J}_{B}$ with
increasing $N$ can be seen. 
Although experimental moment of inertia exhibits a gap between  
Figs.~\ref{fig:MOI}(a) and (b) (from $N=$88 to 90), 
neither ${\cal J}_{B}$ nor ${\cal J}_{F}$ does not 
follow this trend, showing only gradual changes.  This could be 
due to the absence of the particle number conservation in the Skyrme
EDF calculation.  We do not touch on this point in this paper. 

Figures~\ref{fig:MOI}(d) and (e) show, respectively, the
derived $\alpha$ value for $^{146-150}$Sm and $^{152-158}$Sm.
First we notice an overall trend that $\alpha$ does not 
change so much, while the IB value of ${\cal J}_{F}$ changes by 
an order of magnitude.  Although the $\alpha$ values for 
$^{146-150}$Sm do not make much sense, this is of certain 
interest.  

Figure~\ref{fig:MOI}(c) shows the moments of inertia for 
$^{232-236}$U, which are rather flat.  We point out that 
the calculated moment of inertia, ${\cal J}_{F}$=${\cal J}_{B}$ 
with the LL term, turned out to be about twice large as that of 
$^{152-158}$Sm.  This dramatic change is consistent with 
experiment, although somewhat overshoots experimental changes.

\begin{figure}
 \includegraphics[width=7.5cm]{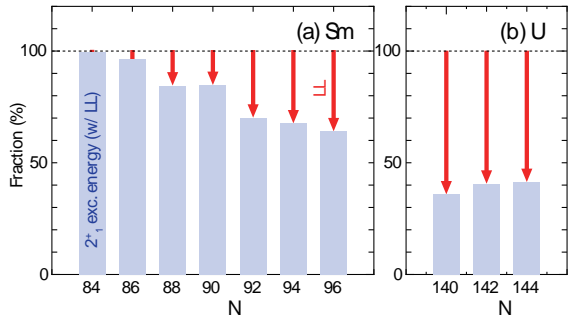}
 \caption{(Color online) Fraction of $E_{\rm x}(2^{+}_{1})$ 
 with the LL term, normalized with respect to $E_{\rm x}(2^{+}_{1})$
 without the LL term, for (a) Sm and (b) U isotopes. Arrows represent
 the LL matrix elements normalized by $E_{\rm x}(2^{+}_{1})$ (w/o LL). } 
\label{fig:LL}
\end{figure}

We shall discuss eigenvalues of $H_{B}^{\prime}$ in eq.~(\ref{eq:bh}) 
obtained by the diagonalization by NPBOS code \cite{npbos}. 
We first investigate to what extent $E_{\rm x}(2^{+}_{1})$ is lowered 
by the LL term. 
Figure~\ref{fig:LL} shows the fraction of this lowering, by normalizing 
it with respect to the $E_{\rm x}(2^{+}_{1})$ without the LL term, 
for (a) Sm and (b) U isotopes. 
This lowering is, as indicated by the arrows in Fig.~\ref{fig:LL},   
$>$30\% for $^{92\sim96}$Sm and $>$60\% for $^{140\sim144}$U. 
On the other side, it is almost vanished or quite small for 
$N=84 \sim 90$.  Thus, it may not affect the IBM description much,
even if one keeps the LL term in all nuclei.  We do not take it, 
because the present derivation does not give physical basis 
for the LL term for nuclei without strong deformation.

\begin{figure}
 \includegraphics[width=7.0cm]{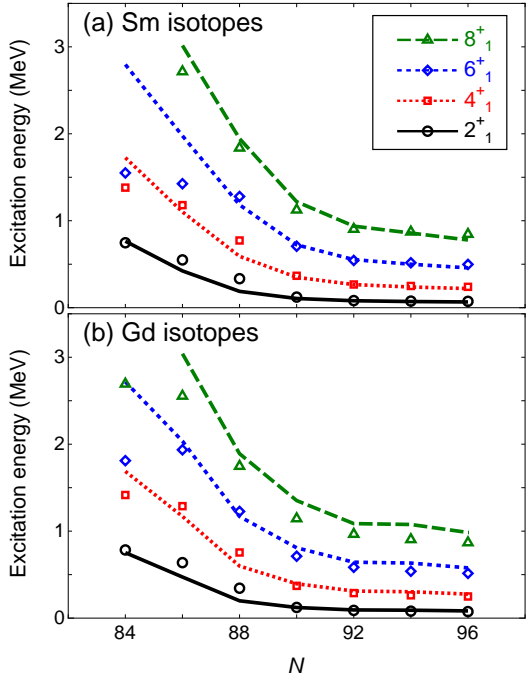}
 \caption{(color online) 
Experimental \cite{data} (symbols) and calculated (curves) yrast spectra
 for (a) Sm and (b) Gd isotopes. }
\label{fig:spectra}
\end{figure}

Figure~\ref{fig:spectra} shows the evolution of low-lying yrast
spectra for (a) Sm and the neighboring (b) Gd
isotopes as functions of $N$. 
For both Sm and Gd isotopes, the LL term is included for
$N\geqslant 90$, but is not included for $N\leqslant 88$, based 
on the criterion discussed above. 
The IBM parameters for Gd isotopes are derived similarly to 
those used for Sm isotopes. 
Figures.~\ref{fig:spectra}(a) and \ref{fig:spectra}(b) indicate that 
calculated spectra become more compressed with $N$
and exhibit rotational feature for $N\geqslant 90$, similarly to the
experimental trends \cite{data}. 
%
One notices a certain deviation at $N=88$, where the Skyrme PES 
favors stronger deformation and the calculated excitation energies
are somewhat too low \cite{nsofull}.  

\begin{figure}[ctb]
 \includegraphics[width=7.0cm]{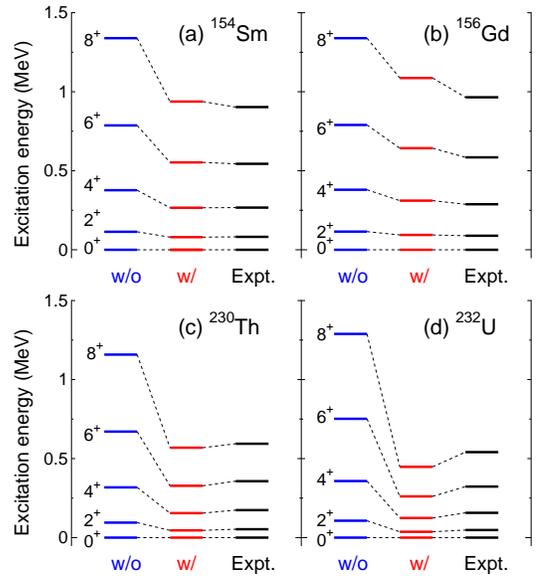}
 \caption{(Color online) 
Level schemes of (a) $^{154}$Sm, (b) $^{156}$Gd, (c) $^{230}$Th and (d)
 $^{232}$U nuclei. 
Calculated spectra with (w/) and without (w/o) the LL
 term, are compared with experimental \cite{data} spectra. }
\label{fig:level}
\end{figure}

Figure~\ref{fig:level} shows yrast levels of $^{154}$Sm, $^{156}$Gd,
$^{230}$Th and $^{230}$U nuclei as representatives of rotational nuclei. 
The LL term is included for these nuclei, as they fulfill the criterion.  
For $^{230}$Th, the parameters of $H_{B}$ take almost the same values as 
those for the $^{232}$U nucleus. 
A nice overall agreement arises between the theoretical 
and the experimental \cite{data} spectra, and the contribution of  
the LL term to it is remarkable.
Particularly for $^{154}$Sm and $^{230}$Th, the calculated spectra look 
nearly identical to the experimental ones. 

We comment on side-band levels. 
The deviations of $\beta$-bandhead ($0^{+}_{2}$) and
$\gamma$-bandhead ($2^{+}_{2}$) energies are improved by tens of keV 
by the LL term.  
However, these band-head energies are still much higher than experimental 
ones.  Thus, there are still open questions on side-band levels. 
On the other hand, the relative spacing inside the bands is reduced by 
hundreds of keV, producing certain improvements. 

We mention some studies deriving a collective Hamiltonian from 
a given EDF where the mean-field PES supplemented with zero point rotational 
and vibrational corrections are treated as an effective potential
\cite{CollSk,CollGo,CollRHB}. 
A generalized kinetic energy term emerges in such approaches. 
In the present work, we compare the results of Skyrme EDF with the
corresponding results of the mapped boson system, at the levels of the PES 
and the rotational response. 
The kinetic energies of nucleons are included in both levels, 
while the rotational kinetic-like boson term appears from the latter.

In summary, we have proposed a novel formulation of the IBM
for rotational nuclei.   
The rotation of strongly deformed multi-nucleon system differs, in its 
response to the rotational cranking, from its boson image obtained 
by the mapping method of
Ref.~\cite{nso} where the PES at rest is considered.   
Significant differences then appear in moment of inertia between
nucleon and boson systems. We have shown that this problem is remedied 
by introducing the LL term into the IBM Hamiltonian.
The effect of the LL term  makes essential contribution to 
rotational spectra, solving the longstanding problem of too small 
moment of inertia microscopically.    
Experimental data are reproduced quite well, 
without any phenomenological adjustment.  
The mapping of Ref.~\cite{nso} appears quite sufficient for 
vibrational and $\gamma$-unstable nuclei, and the present 
study makes the IBM description of strongly deformed nuclei  
sensible theoretically and empirically.  Thus, we seem to have
come to the stage of having microscopic basis of the IBM in 
all situation at the lowest order.  On the other hand, this 
achievement is partly due to the successful description of Skyrme 
mean-field model.  
The feature discussed in this paper is related to the question as to
whether the IBM can be applied to deformed nuclei or not \cite{BMc}.  
The present work indicates that the rotational response is substantially
different between fermions and bosons, but the difference can be
incorporated into the IBM in a microscopic way. 

The authors acknowledge Dr. A.~Gelberg and Dr. D.~Vretenar for valuable 
comments. 
This work was supported in part by grants-in-aid for Scientific
Research (A) 20244022 and No.~217368, and by the JSPS Core-to-Core
program EFES. 
Authors K.N. and L.G. have been supported by the JSPS Fellowship
program.

\end{document}